\documentclass[5p]{elsarticle}

\usepackage{graphicx}

\usepackage{amssymb}

\def\kevc1{\ifmmode\mathrm{\ keV/{\mit c}}
          \else$\mathrm{\ keV/{\mit c}}$\fi}
\def\MeVc1{\ifmmode\mathrm{\ MeV/{\mit c}}
          \else$\mathrm{\ MeV/{\mit c}}$\fi}
\def\mevc1{\ifmmode\mathrm{\ MeV/{\mit c}}
          \else$\mathrm{\ MeV/{\mit c}}$\fi}
\def\gevc1{\ifmmode\mathrm{\ GeV/{\mit c}}
          \else$\mathrm{\ GeV/{\mit c}}$\fi}
\def\kevc2{\ifmmode\mathrm{\ keV/{\mit c}^2}
          \else$\mathrm{\ keV/{\mit c}^2}$\fi}
\def\Mevc2{\ifmmode\mathrm{\ MeV/{\mit c}^2}
          \else$\mathrm{\ MeV/{\mit c}^2}$\fi}
\def\Gevc2{\ifmmode\mathrm{\ GeV/{\mit c}^2}
          \else$\mathrm{\ GeV/{\mit c}^2}$\fi}
\def\Gev2c2{\ifmmode\mathrm{\ GeV^2/{\mit c}^2}
          \else$\mathrm{\ GeV^2/{\mit c}^2}$\fi}
\def\Pp{\ifmmode\mathrm{p}
         \else$\mathrm{p}$\fi}
\def\Pap{\ifmmode\mathrm{\overline{p}}
         \else$\mathrm{\overline{p}}$\fi}
\def\PgL{\ifmmode\mathrm{\Lambda}
          \else$\mathrm{\Lambda}$\fi}
\def\PagL{\ifmmode\mathrm{\overline{\Lambda}}
            \else$\mathrm{\overline{\Lambda}}$\fi}
\def\PgS{\ifmmode\mathrm{\Sigma^-}
          \else$\mathrm{\Sigma^-}$\fi}
\def\PagS{\ifmmode\mathrm{\overline{\Sigma}^+}
            \else$\mathrm{\overline{\Sigma}^+}$\fi}
\def\PgX{\ifmmode\mathrm{\Xi}
          \else$\mathrm{\Xi}$\fi}
\def\PagX{\ifmmode\mathrm{\overline{\Xi}}
            \else$\mathrm{\overline{\Xi}}$\fi}
\def\PgLc{\ifmmode\mathrm{\Lambda_c}
          \else$\mathrm{\Lambda_c}$\fi}
\def\PagLc{\ifmmode\mathrm{\overline{\Lambda}_c}
            \else$\mathrm{\overline{\Lambda}_c}$\fi}
\begin{document}

\begin{frontmatter}



\title{Exploring the potential of antihyperons in nuclei with antiprotons}

\author{J.~Pochodzalla}
\ead{pochodza@kph.uni-mainz.de}
\address{Johannes Gutenberg-Universit{\"a}t Mainz, Institut f{\"u}r Kernphysik, D-55099 Germany}

\begin{abstract}

A simple method to explore the interaction of antihyperons in nuclei
by exclusive hyperon-antihyperon pair production close to threshold
in antiproton nucleus interactions is proposed. Due to energy and
momentum conservation event-by-event transverse momentum
correlations of the produced hyperons and antihyperons contain
information on the difference between their potentials. A schematic
Monte Carlo simulation is used to illustrate the sensitivities of
the proposed method for the reaction 1.66\gevc1 \Pap$^{12}$C
$\rightarrow$ {\PgL\PagL}. For produced D-meson pairs at 6.7\gevc1
the sensitivity of the transverse momenta correlation will probably
be too small to deduce differences between the potentials for D$^+$
and D$^-$ mesons. However, for {\PgX\PagX} pairs produced at
2.9\gevc1 the asymmetry is sufficiently sensitive to predicted
differences between the {\PgX} and {\PagX} potentials.
\end{abstract}
\end{frontmatter}


Based on G-parity transformation~\cite{Lee56} D\"urr and Teller
predicted within an early form of a relativistic field theory a
strongly attractive potential for antiprotons in
nuclei~\cite{Duer56a,Duer56b}. It is however obvious that G-parity
transformation can provide a link between the $NN$ and
$N\overline{N}$ interaction at most for distances where meson
exchange is a valid concept~\cite{Dov80,Fae82}. For distances lower
than about 1 fm, quark degrees of freedom may play a decisive role.
The study of the potential of antibaryons in nuclei may therefore
help to elucidate the role of the quark-gluon structure of baryons
for the short-range baryon-baryon force.

Early studies of antiproton-nucleus scattering cross
sections~\cite{Duer58,Gol58b} showed however disagreement with such
a strong attractive potential. Later, X-ray transitions in
antiprotonic atoms~\cite{Bar72,Bac72,Rob77,Pot78} (an overview on
subsequent experimental studies can be found in Ref. \cite{Fri07})
gave also hints for an attractive potential albeit with large
uncertainties~\cite{Aue81,Bat81}. Other analyses favor shallow real
and deep imaginary potentials (for example \cite{Bon86}). More
comprehensive studies \cite{Fri05} of antiprotonic X-rays  as well
as recent analyses of the production of antiprotons in reactions
with heavy ions resulted in real attractive potentials in the range
of about -100 to -150~MeV~\cite{Teis94,Spie96,Sib98}.

Concerning baryons beyond SU(2), only for \PgL\ hyperons reliable
information on their nuclear potential is available from hypernuclei
studies. No experimental information on the nuclear potential of
antihyperons exists so far. Mishustin and co-workers recently
suggested to study deeply bound antibaryonic nuclei via various
characteristic signals in their decay process~\cite{Mish05,Lar08},
like the production of multi-quark-antiquark clusters,
multifragmentation events with strong radial flow or sharp lines in
meson spectra due to transitions from the Fermi to the Dirac sea.
From the experimental point of view it is however not obvious
whether these proposed observables will provide unique and
quantitative signals of deeply bound antibaryonic systems.

\begin{figure}[t]
\begin{center}
\includegraphics[width=0.6\linewidth]{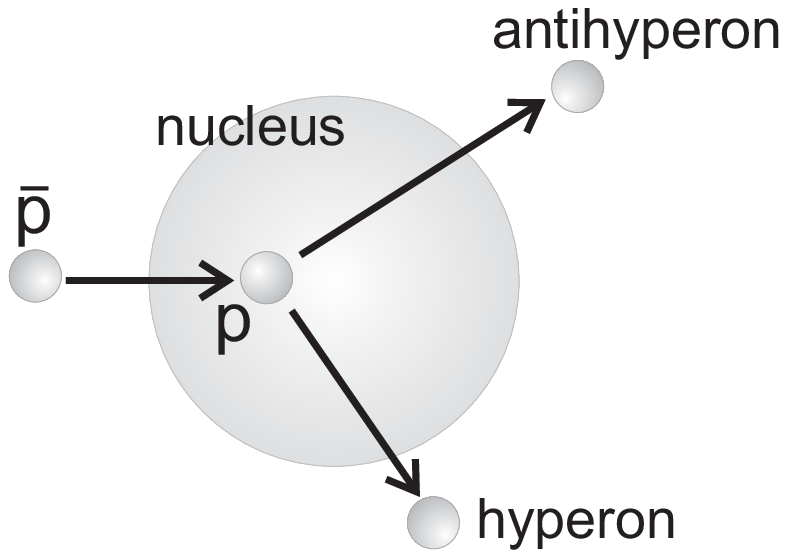}\\
\includegraphics[width=0.9\linewidth]{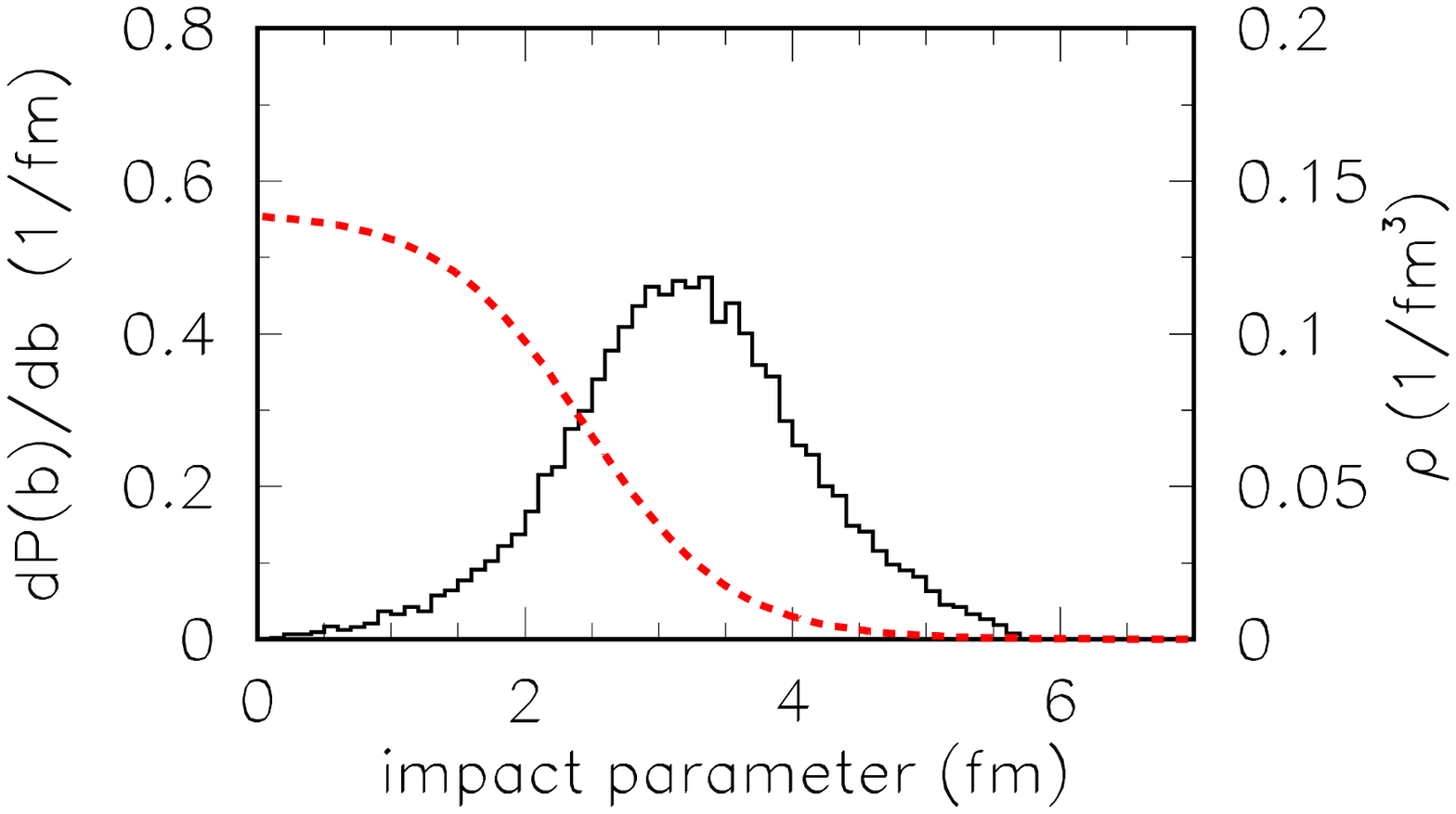}
\end{center}

\caption{Scheme of the reaction proposed to explore the nuclear
potentials of the baryon and the antibaryon (top). The lower part
shows the probability density distribution of the impact parameter
leading to the observation of a \PgL\PagL\ pair in 1.66\gevc1\ \Pap
$^{12}$C$ \rightarrow$ \PgL\PagL\ reactions. In these simulations
default parameters as discussed in the text have been used. For
orientation the dashed line gives the assumed radial density of the
$^{12}$C nucleus (right scale).} \label{fig:01}
\end{figure}

In this letter we show that quantitative information on the
antihyperon potentials relative to that of the corresponding hyperon
may be obtained via exclusive antihyper\-on-hyperon pairs production
close to threshold after an antipro\-ton-proton annihilation within
a complex nucleus (Fig.~\ref{fig:01}). Once these hyperons leave the
nucleus and are detected, their asymptotic momentum distributions
will reflect the depth of the respective potentials. A deep
potential for one species could result in a momentum distribution of
antihyperons which differs from that of the coincident hyperon. This
situation is in line with the case of antiprotons produced in heavy
ion collisions close to threshold~\cite{Teis94,Spie96,Sib98}. The
advantage here is, that we are dealing with a quasi stationary
system having a reasonably well defined geometry and that the
kinematics is determined essentially by energy and momentum
conservation of a (nearly) two-body reaction. However, since in the
{\Pap\Pp} center-of-mass the distribution of the produced
baryon-antibaryon pair will usually not be isotropic, the analysis
can rely only on the {\em transverse} momenta of the outgoing
baryons: because the initial average transverse momentum is equal to
zero and neglecting for the moment rescattering and the Fermi motion
of the struck proton (both effects will be discussed below) the
magnitude of the transverse momenta of the produced baryons and
antibaryons will be equal unless there is a difference in the
effective potentials.

\begin{table*}
\caption{Scalar and vector potentials, S and V, used in the model
calculations. The first three columns give the default values for
\Pp\, \Pap\ and \PgL\ hyperons used in the parameter scans which are
presented in Fig.~\ref{fig:02} and Fig.~\ref{fig:03}. The values in
the last 8 columns were adopted from Ref.~\cite{Tsu03,Sai05} and are
used in the calculations shown in Fig.~\ref{fig:04}.}
\begin{center}
\begin{tabular}{c|rrr|rrrrrrrr}
\hline
potential & \Pp & \Pap & \PgL & \Pp& \Pap& \PgL & \PagL & \PgX  & \PagX  & D$^+$  & D$^-$ \\
\hline

V [MeV]&  300 & 200 & 200 & 125 & -125 &  84 & -84 & 42 & -42 & -42 & 42\\
S [MeV]&  -342 & -342 & -228 &-184 & -184 &-123 &-123 &-61 & -61 & -61 &-61\\
V+S [MeV]& -42 & -142 & -28& -59 & -309 & -39 &-207 &-19 &-103 &-103
&-19\\ \hline
\end{tabular}
\label{tab:01}
\end{center}
\end{table*}

In the following we explore the influence of the potentials on the
transverse momentum distributions of the coincident hyperons and
antihyperons as well as on their event-by-event correlations by
means of a schematic Monte Carlo simulation. Albeit crude, this
classical approach allows to explore the role of different features
of the reaction in a transparent way. As an example we consider the
\Pap$^{12}$C $\rightarrow$ \PgL\PagL\ reaction at {1.66\gevc1},
where existing data \cite{Barnes91,Pomp} demonstrate the feasibility
of such measurements. Since the method relies essentially on
momentum and energy conservation, an extension to other
hadron-antihadron pairs produced exclusively in antiproton-nucleus
collisions close to their respective thresholds is straight forward.
In future, such reactions can be studied at the international
Facility for Antiproton and Ion Research {\sc FAIR} \cite{FAIR08}
with e.g. the planned ${\overline{\rm P}}$ANDA experiment
\cite{Panda}.

\begin{figure*}[t]
\begin{center}
\includegraphics[width=0.97\linewidth]{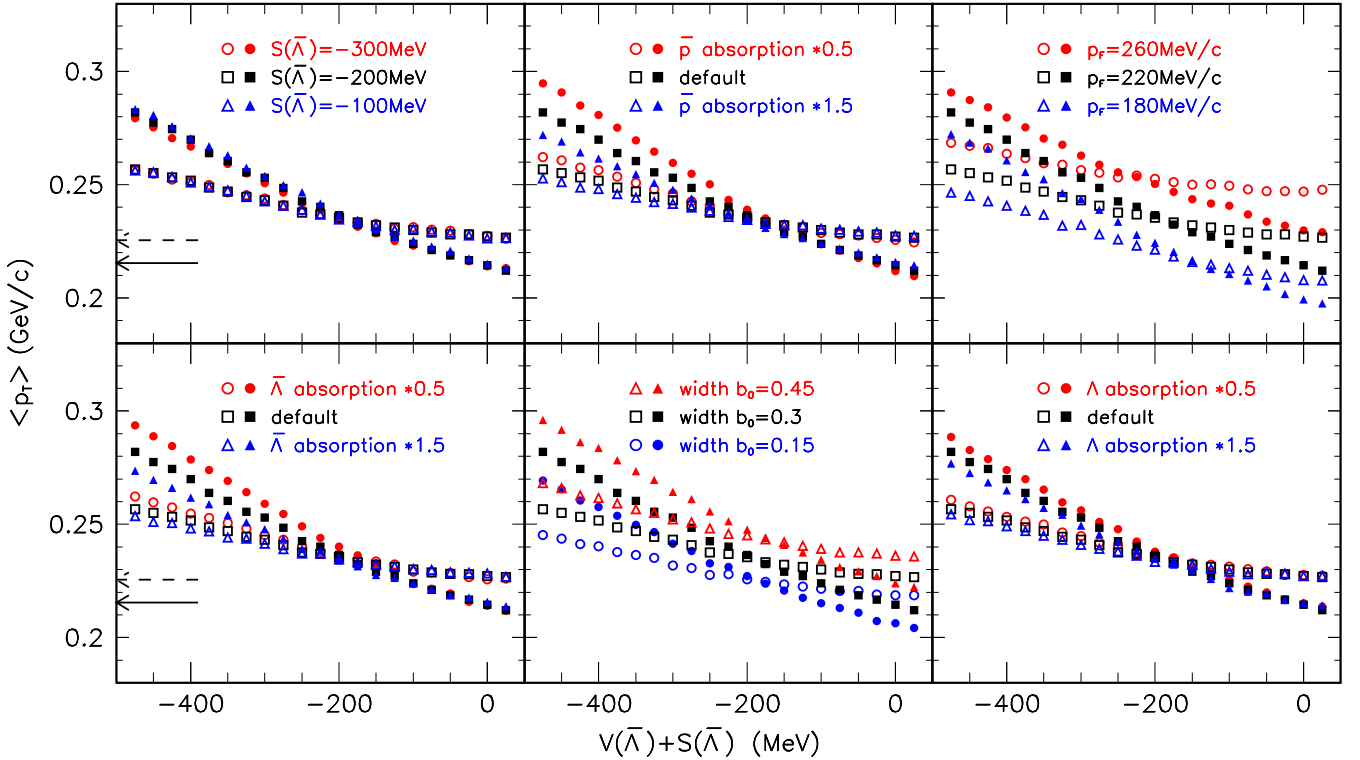}
\end{center}
\caption{Transverse momentum of \PagL\ hyperons (open symbols) and
\PgL\ hyperons as a function of the total \PagL\ potential for
different parameter sets. The dashed and solid arrows mark the
result for {\PagL} and {\PgL} if all individual \PgL\ and \PagL\
potentials are set to 0. } \label{fig:02}
\end{figure*}

The absorption of the antiprotons entering the target nucleus
determines the points of annihilation inside the nucleus and the
paths which the eventually produced hyperons and antihyperons have
to pass inside the nucleus prior to emission. For the proton density
we adopted a Fermi-type distribution
\begin{equation}
\rho(r)=\frac{\rho_0} {1+exp((r-r_0{\cdot}A^{1/3})/d_0)},
\label{eq:01}
\end{equation}
where $r$ denotes the radial distance from the center of the target
nucleus and $A$ its mass number. For the radius parameter $r_0$ and
the surface diffuseness $d_0$ default values of 1.07~fm and 0.54~fm
were used \cite{Hah56}. Since in the following we focus on light
nuclei, equal neutron and proton density distributions were assumed.

The initial \Pap\Pp\ annihilation is controlled by an {\Pap}N
annihilation cross section of 50~mb~\cite{PDG}. The lower part of
Fig.~\ref{fig:01} shows the probability density distribution of the
impact parameter leading to the emission of a \PgL\PagL\ pair in
1.66\gevc1 \Pap$^{12}$C $\rightarrow$ \PgL\PagL\ reactions. For
orientation the dashed line gives the assumed radial density profile
of the $^{12}$C nucleus. Because of the strong absorption of the
antihyperons, the {\em emitted} hyperon-antihyperon pairs are -
unlike in inclusive reactions \cite{Sibirtsev99,Lenske05} - created
close to the corona of the target nucleus at an average impact
parameter of 3.1~fm and a typical density of 20 to 25\% of the
central nuclear density.

For both, the emitted {\PgL}'s and the {\PagL}'s the inverse of the
average integrated path weighted with the local density along the
path $\langle \int\rho ds \rangle^{-1}$ varies for the parameter
range discussed in this paper in the range from about 800 to 1600
mb. For our default parameters this value is about 1000mb and thus
significantly larger than the typical elastic cross sections in the
relevant momentum range of $<$200mb. As a consequence re-scattering
effects are expected to be small and have been neglected in our
model. Experimentally, re-scattering effects with momentum transfers
beyond the typical Fermi momentum can possibly be reduced by
constraining the azimuthal angle between the hadron and antihadron
momentum.

In reactions close to threshold the Fermi motion of the protons
inside the nuclear target contributes significantly to the final
momenta. Hence the initial proton momentum was sampled from a
distribution
\begin{equation}
dP(p,\theta,\phi) \propto (1+e^{(E-E_F)/kT})^{-1}p^2 \sin(\theta)
{dp~d\theta~d\phi}~~ \label{eq:02}
\end{equation}
where a default Fermi energy $E_F$ corresponding to a momentum of
$p_F$=~220{\MeVc1} was used~\cite{Moniz71}. Quasifree meson
scattering experiments suggest \cite{Wise93,Kormanyos95,Fujii01}
that owing to the fact that the \Pap\Pp\ annihilations happens in
the periphery of the target nucleus at subsaturation density,
significantly lower Fermi momenta may be expected. We therefore
varied $p_F$ in the range of 180 to 260{\MeVc1}. For the diffuseness
parameter a value of $T$=~1~MeV was used and the maximum possible
energy was determined by $E_F+E_B$, where $E_B$=8~MeV denotes the
typical nucleon binding energy.

Antilambdas produced in \Pap\Pp\ annihilations are emitted
preferentially in the direction of the incident antiproton
\cite{Barnes91}. For the 1.66\gevc1 {\Pap}$^{12}$C $\rightarrow$
\PgL\PagL\ reaction \cite{Barnes91,Tabakin91,Barnes96} the
probability distribution of the center of mass angle $\theta_{cm}$
of the outgoing antihyperon can be described by:
\begin{equation}
dP(\theta_{cm})\propto \frac{a_0 +
\exp((\cos\theta_{cm}-1)/b_0)}{1+a_0} {d(\cos\theta_{cm})}
\label{eq:03}
\end{equation}
with default values for the constant term $a_0$ and the width $b_0$
of 0.1 and 0.3, respectively.

Lacking any detailed experimental information it is plausible to
assume that the annihilation cross sections for antihyperons show a
similar momentum dependence as the \Pap\Pp\ system \cite{Weber02}.
We therefore parameterized the absorption cross section of the
${\PagL}$ as
\begin{equation}
\sigma^{ann}_{{\PagL}N}=100mb/(p_{{\PagL}}+1)
 \label{eq:04}
\end{equation}
with the ${\PagL}$ momentum given in {\gevc1}. For the \PgL\
hyperons a momentum independent inelastic cross section of 20~mb was
adopted.

The energy and the momentum of the baryons propagating within the
nucleus are related according to \cite{Yamazaki99}:
\begin{equation}
(E-V)^2=(M_0+S)^2+\mathbf{P_{in}}^2
\label{eq:05}
\end{equation}
Here $V$ and $S$ denote the real part of the vector and scalar
potential, respectively. The relation between the momenta inside and
outside of the nuclear potential are approximated by
\begin{equation}
\mathbf{P_{out}}^2 + M_0^2=(\sqrt{(M_0+S)^2+\mathbf{P_{in}}^2}+V)^2.
\label{eq:06}
\end{equation}
Refractive effects at the potential boundary were ignored. For
simplicity no momentum dependence of these potentials was considered
in our schematic simulation. The default parameters for the scaler
and vector potentials of the various baryons at normal nuclear
density $\rho_0$ are listed in Tab.~\ref{tab:01}. For the proton and
the \PgL\ hyperon (given by 2/3 of that of the proton) these values
give rise to typical total potentials of -42 and -28~MeV,
respectively. For the antiproton the summed potential is in the
range of the more recent experimental
results~\cite{Fri05,Teis94,Spie96,Sib98}.

Since the antiproton annihilation and the subsequent \PagL\PgL-pair
production take place in the nuclear periphery at low densities
$\rho$ (see Fig.~\ref{fig:01}), the local potentials are expected to
be reduced. We assumed for simplicity a linear density dependence
$\propto \rho/\rho_0$ for all vector and scalar potentials. All
numbers for potentials quoted below refer to the value at normal
nuclear density $\rho_0$.

In a last step a finite momentum resolution of 10\% was applied to
mimic possible experimental uncertainties.

Fig.~\ref{fig:02} shows the average transverse momenta of \PgL\
hyperons (closed symbols) and \PagL\ hyperons (open symbols) as a
function of the total \PagL\ potential $V(\PagL)+S(\PagL)$ for
various parameter sets. In all plots the black points result from
the default parameter set. The transverse momenta of the \PagL\
hyperons drop with decreasing depth of the total \PagL\ potential.
For the \PgL\ hyperons this drop is even more pronounced. This
surprising behavior can be traced back to the imposed momentum
conservation and the different sign of the vector potentials for
hyperons and antihyperons. Thus within our schematic model an
agreement between the transverse momenta of hyperons and
antihyperons would not necessarily imply that both encounter
identical potentials.

Even if all antihyperon and hyperon potentials are set to zero one
finds different average transverse momenta for {\PagL}'s and
{\PgL}'s of 225 and 215{\MeVc1}, respectively. They are marked by
the dashed and solid arrows in Fig.~\ref{eq:02}. This difference is
caused by the anisotropy in the $\theta_{cm}$-distribution (Eq.
\ref{eq:03}) and the relativistic transformation of the isotropic
Fermi momentum. We checked that if either the assumed Fermi momentum
distribution in the target is switched off or if an isotropic
$\theta_{cm}$-distribution is assumed one obtains equal average
transverse momenta for {\PgL} and {\PagL} hyperons. The largest
sensitivities to variations of the model parameters are observed for
the assumed Fermi momentum (left middle panel) and the assumed
anisotropy (left lower panel). For all other parameters the
sensitivity of $p_T$ is significantly weaker. Thus, the simultaneous
measurement of transverse momenta of hyperons and antihyperons can
be used to adjust the Fermi momentum in the calculations.

\begin{figure}[t]
\includegraphics[width=0.91\linewidth]{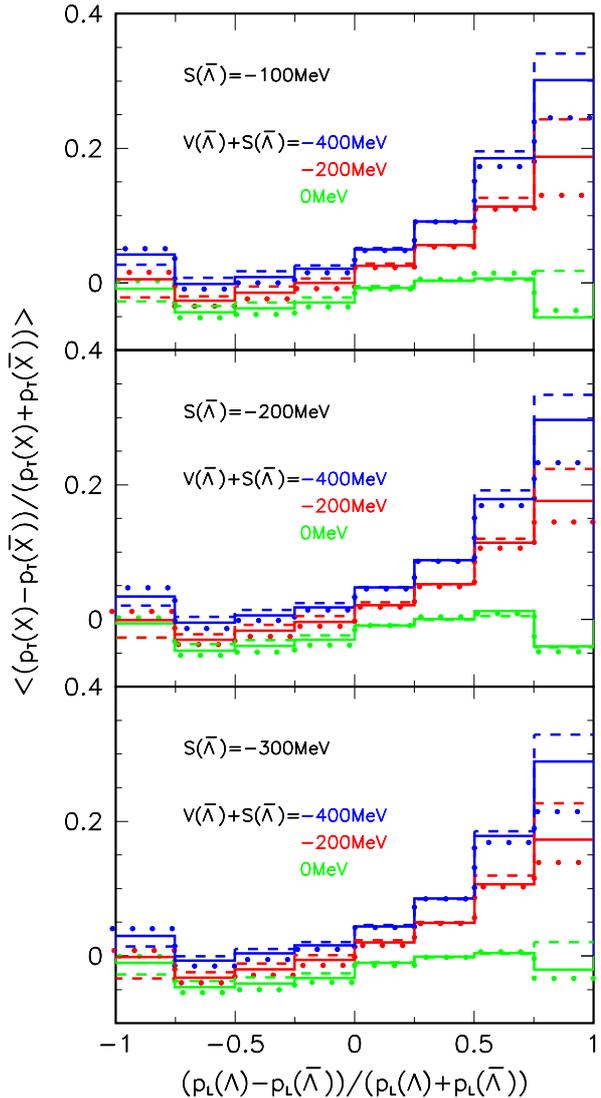}

\caption{Average transverse momentum asymmetry as a function of the
longitudinal momentum asymmetry for different parameter pairs of the
scaler and vector \PagL\ potentials. In each panel calculations with
3 different Fermi momenta of 180\MeVc1\ (dashed lines), 220\MeVc1\
(solid lines), and 260\MeVc1\ (dotted lines) are overlaid.}
\label{fig:03}
\end{figure}

\begin{figure}[t]
\includegraphics[width=0.95\linewidth]{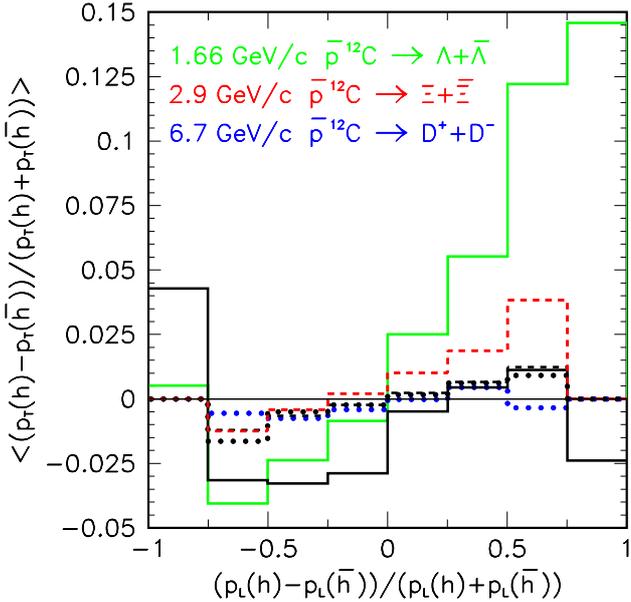}

\caption{Average transverse momentum asymmetry of {\PgL\PagL}(solid
green line), \PgX\PagX\ (red dashed line) and D$^+$D$^-$ pairs (blue
dotted line) produced exclusively in 1.66, 2.9 and 6.7\gevc1\ \Pap
$^{12}$C interactions, respectively. In these calculations the
potentials given in the last eight columns of Table~\ref{tab:01}
were used. The black solid, dashed and dotted histograms are the
result if all scalar and vector potentials of the produced
{\PgL\PagL}, {\PgX\PagX} and D$^+$D$^-$ pairs are set to 0.
Asymmetries identical to 0 in the first and last bins signal zero
counts.} \label{fig:04}
\end{figure}

Studying only the average transverse momentum distributions
separately does obviously not allow to extract unambiguous
information on the potential of antihyperons. On the other hand, a
difference between transverse momenta of the concident hadron and
antihadron within one event reflects directly the different
potentials. In order to study this correlation and to reduce the
influence of the center of mass angle $\theta_{cm}$ we suggest to
explore the transverse momentum asymmetry $\alpha_{T}$ as a function
of the longitudinal asymmetry $\alpha_{L}$. Here, $\alpha_{T}$ and
$\alpha_{L}$ are defined for each event in terms of the transverse
and longitudinal momenta, respectively:

\begin{equation}
\noindent
\alpha_{T}=\frac{p_{T}(\PgL)-p_{T}(\PagL)}{p_{T}(\PgL)+p_{T}(\PagL)}
~~~,~~
\alpha_{L}=\frac{p_{L}(\PgL)-p_{L}(\PagL)}{p_{L}(\PgL)+p_{L}(\PagL)}
.~~ \label{eq:07}
\end{equation}
The Fermi motion of the struck proton inside the target may provide
a total transverse momentum which will be different for each event.
However, in case the scalar and vector potentials of hyperons and
antihyperons are equal, the average $\langle \alpha_T\rangle$ is
expected to be 0 except for small asymmetries caused by the combined
effect of the Fermi motion, the anisotropic angular distribution and
the different absorption cross sections for {\PagL}'s and {\PgL}'s.

The histograms in fig.~\ref{fig:03} show the average transverse
asymmetry $\langle\alpha_T\rangle$ for various bins in $\alpha_L$.
Calculations were done for three different scalar potential
$S({\PagL})$ of -100, -200, and -300~MeV, respectively. In each
panel, the different colored histograms are the results for total
potentials of $V(\PagL)+S(\PagL)$=0, -200 and -400~MeV,
respectively. Furthermore, in each plot calculations with three
different Fermi momenta of 180, 220, and 260{\MeVc1} are overlaid.
For all other parameters the default values were used. Variations in
the first and last bins are partly caused by the low number of
events in these bins resulting in statistical errors $\delta\alpha_T
\simeq$ 0.01 and 0.02, respectively. While the average transverse
momenta are very sensitive to the choice of the Fermi momentum, the
transverse momentum asymmetry is not. Using different
parametrizations of the Fermi motion (based e.g. on a local density
approximation) gave rather similar final results provided the Fermi
momentum parameter was tuned to similar average transverse hyperon
and antihyperon momenta. Indeed, for a given \PgL\ potential,
$\alpha_T$ is mainly determined by the total potential
$V(\PagL)+S(\PagL)$ as indicated by the overlap of the histograms of
the same color. At negative values of $\alpha_L$ the ${\PagL}$
momenta are relative large and consequently the sensitivity of
$\alpha_T$ to the potential is weaker.

The average transverse asymmetry is non-zero even if the total
potential $V({\PagL})+S({\PagL})$=0 (green histograms). A very
similar behavior is found even if all scalar and vector potentials
for \PagL\ and \PgL\ are set to 0. Like in the case of the average
$p_T$, this is caused by the interplay between the isotropic Fermi
motion and the anisotropic cm-distribution and - although less
important - by the different absorption cross section. Also the
assumed scaling of the momentum resolution with the absolute
momentum causes a small positiv correlation between $\alpha_T$ and
$\alpha_L$. We also checked that the results are rather robust
despite significant changes - by typically $\pm$50\% - of all other
parameters like absorption cross sections and (anti)proton
potentials. Systematic relative shifts remained usually below
$\delta\alpha/\alpha=\pm0.15$. Furthermore, neglecting the momentum
dependence of the {\PagL} absorption and assuming constant
absorption cross sections of 100mb and 20mb yields asymmetries very
similar to the momentum dependent cross section of Eq. \ref{eq:04}
scaled by a factor 1.5 and 0.5, respectively.

As already mentioned before, this kinematic method of transverse
momentum correlations can in principle be applied to each
hadron-antihadron pair produced exclusively in $\overline{p}A$
interactions. As an example, the colored solid, dashed and dotted
histograms in figure \ref{fig:04} show the average transverse
momentum asymmetry of {\PgL\PagL}, {\PgX\PagX} and D$^+$D$^-$ pairs
produced in 1.66, 2.9 and 6.7\gevc1\ \Pap\ + $^{12}$C interactions,
respectively. For simplicity, isotropic center-of-mass distributions
were assumed in case of the {\PgX\PagX} and D$^+$D$^-$ production.
For the \PgX\ and \PagX\ baryons the same absorption cross sections
as for the \PgL\ and \PagL\ were adopted, whereas for D$^-$ and
D$^+$ mesons energy independent absorption cross sections of 10 and
90mb, respectively, were taken. The scalar and vector potentials
were inspired by Refs.~\cite{Tsu03,Sai05} and are listed in
Tab.~\ref{tab:01}. To mimic experimental effects a relative
resolution for the momentum reconstruction of 5\% was also taken
into account.

For orientation, the black histograms in Fig.~\ref{fig:04} show the
asymmetries if all scalar and vector potentials of the outgoing
hadrons and antihadrons are set to 0. While for {\PgX\PagX} (dashed)
and D$^+$D$^-$ pairs (dotted) these histograms are symmetric around
$\alpha_L$=0, it is not the case for {\PgL\PagL} pairs (solid line).
As already mentioned before, this is caused by the assumed
anisotropic c.m.-distribution of the {\PgL\PagL} production.

Also these calculations confirm the robustness of the transverse
momentum asymmetry with respect to variations of the model
parameters: as expected from the similar values for S and V
(Tab.~\ref{tab:01}), the asymmetries for the {\PgL\PagL} pairs
(green histogram in Fig.~\ref{fig:04}) are indeed close to the red
histograms shown in Fig.~\ref{fig:03}. We also would like to note
that a quite similar result is found for {\PgS\PagS} pairs. In a
purely classical, non-relativistic picture the asymmetry is of the
order of ${\Delta}U/4\cdot E_0$, where ${\Delta}U$ is the potential
difference and $E_0$ the typical kinetic energy of the hadrons. In
line with this consideration the large laboratory momenta of the
\PgX\ hyperons and the D mesons explain the smaller asymmetries for
the heavier particles.

To demonstrate the experimental feasibility of the proposed
measurement one may consider as an example the bins of 0.25$\leq
\alpha_T<$ 0.5 and 0.5$\leq \alpha_T<$ 0.75 where a sizable
asymmetry is predicted. Depending on the choice of parameters these
bins contain approximately 6-10\% and 1-3\%, respectively, of the
total number of events. The $\alpha_T$-distributions have a typical
width of 0.3. At the expected {\PgL\PagL} detection rates at
${\overline{\rm P}}$ANDA \cite{PANDA_PB} measurement periods of a
few minutes will be sufficient to reach a relative statistical
uncertainty of better than 10\% for $\alpha_T$ within these two
bins. Given the relative large cross section for
{\Pp\Pap}$\rightarrow$ {\PgX\PagX} at 2.9 \gevc1\ of ${\sim}1\mu$b
\cite{Kai94} the sensitivity of the transverse asymmetry (dashed
histograms in Fig.~\ref{fig:04}) is sufficiently large to explore
the {\PgX\PagX} pair production at the future {\sc FAIR} facility.
 In case of {\PgX\PagX} pairs a measurement of
 $\alpha_T$ with a precision of 10\% in the same two bins will
 require typically 2 and 10 hours, respectively. For D-meson
pairs, however, the large momenta relative to the target remnant
($>3${\gevc1}) and the low production cross section ($\sim$10nb)
casts a meaningful measurement of the transverse momentum asymmetry
in doubt for the case of the potential difference of
$\approx$50-100~MeV given in Tab.~\ref{tab:01}. (cf. the dotted
black and dotted blue histograms in Fig.~\ref{fig:04}). Only for
significantly deeper potentials than the ones listed in
Tab.~\ref{tab:01} a measurable asymmetry can be expected. But even
then the estimated measurement periods will significantly exceed one
month.

The fact that energy and momentum conservation are the main
ingredient of the proposed method raises hope that similar results
might be obtained by more realistic calculations taking for example
the momentum dependence of the potentials into account. Since most
of the {\em emitted} hyperon-antihyperon pairs are created in the
nuclear periphery at subsaturation density, a neutron skin of
neutron rich target nuclei may help to explore different effective
potentials. Significant deflections at the potential boundary which
are ignored in the present work may be at least partly eliminated by
demanding that the target nucleus remains intact. Furthermore, it
may be interesting to study questions related to e.g. the formation
time \cite{Cas02} by using target nuclei of different size.

The author thanks S. Pomp, T. Johansson and W. Eyrich for helpful
discussions. We acknowledge financial support from the
Bundesministerium f\"ur Bildung und For\-schung (bmb+f) under
contract number 06MZ225I.


\begin{thebibliography}{99}

\bibitem{Lee56}
T.D. Lee and C.N. Yang, Nuovo Cim. {\bf 3}, 749 (1956).

\bibitem{Duer56a}
Hans-Peter D\"urr and Edward Teller, Phys. Rev. {\bf 101}, 494
(1956).

\bibitem{Duer56b}
Hans-Peter D\"urr, Phys. Rev. {\bf 103}, 469 (1956).

\bibitem{Dov80}
C.B. Dover and J.M. Richard, Phys. Rev. C {\bf 21}, 1466 (1980).

\bibitem{Fae82}
A. Faessler, G. L\"ubeck and K. Shimizu, Phys. Rev. D {\bf 26}, 3280
(1982).

\bibitem{Duer58}
Hans-Peter D\"urr, Phys. Rev. {\bf 109}, 1347 (1958).


\bibitem{Gol58b}
G. Goldhaber and J. Sandweiss, Phys. Rev. {\bf 110}, 1476 (1958).

\bibitem{Bar72}
P. D. Barnes, S. Dytman, R. A. Eisenstein, W. C. Lam, J. Miller, R.
B. Sutton, D. A. Jenkins and R. J. Powers, M. Eckhause, J. R. Kane,
B. L. Roberts, R. E. Welsh, A. R. Kunselman, R. P. Redwine and R. E.
Segel, Phys. Rev. Lett. {\bf 29}, 1132 (1972).

\bibitem{Bac72}
G. Backenstoss, A. Bamberger, T. Bunaciu, J. Egger, H. Koch, U.
Lynen, H. G. Ritter, H. A. Schmitt, A. Schwitter, Physics Letters B
{\bf 41}, 552 (1972).


\bibitem{Rob77}
P. Roberson, T. King, R. Kunselman, J. Miller, R.J. Powers, P.D.
Barnes, R.A. Eisenstein, R.B. Sutton, C.R. Cox, M.Eckhause, J.R.
Kane, A.M. Rushton, W.F. Vulcan, and R.E. Welsh, Phys. Rev C {\bf
16}, 1945 (1977).

\bibitem{Pot78}
H. Poth, G. Backenstoss, I. Bergström, P. Blüm, J. Egger, W.
Fetscher, R. Guigas, R. Hagelberg, N. Hassler, C. J. Herrlander, M.
Izycki, H. Koch, A. Nilsson, P. Pavlopoulos, H. P. Povel, K. Rolli,
I. Sick, L. Simons, A. Schwitter, J. Sztarkier, and L. Tauscher,
Nucl. Phys. A {\bf 294}, 435 (1978).


\bibitem{Fri07}
E. Friedman and A. Gal, Phys. Rep. {\bf 452}, 89 (2007).

\bibitem{Aue81}
E. H. Auerbach, C. B. Dover, and S. H. Kahana, Phys. Rev. Lett. {\bf
46}, 702 (1981).

\bibitem{Bat81}
C.J. Batty, Nucl. Phys. A {\bf 372}, 433 (1981).

\bibitem{Bon86}
R. Bonetti and M.S. Hussein, J. Phys. G: Nucl. Phys. {\bf 12}, L119
(1986).

\bibitem{Fri05}
E. Friedman, A. Gal and  J. Mares, Nucl. Phys. A {\bf 761}, 283
(2005).

\bibitem{Teis94}
Stefan Teis, Wolfgang Cassing, Tomoyuki Maruyama and Ulrich Mosel,
Phys. Rev. C {\bf 50}, 388 (1994).

\bibitem{Spie96}
C. Spieles, M. Bleicher, A. Jahns, R. Mattiello, H. Sorge, H.
St\"ocker, and W. Greiner, Phys. Rev. C {\bf 53}, 2011 (1996).

\bibitem{Sib98}
A. Sibirtsev, W. Cassing, G.I. Lykasov and M.V. Rzjanin, Nucl. Phys.
A {\bf 632}, 131 (1998).

\bibitem{Mish05}
I.N. Mishustin, L.M. Satarov, T.J. B\"urvenich, H. St\"ocker, and W.
Greiner, Phys. Rev. C {\bf 71}, 035201 (2005).

\bibitem{Lar08}
A.B. Larionov, I.N. Mishustin, L.M. Satarov, and W. Greiner,
arXiv:0802.1845v2.

\bibitem{Barnes91}
P.D. Barnes, G. Diebold, G. Franklin, C. Maher, B. Quinn, J.Seydoux,
K. Kilian, R. Besold, W. Eyrich, R. v. Frankenberg, A. Hofmann, D.
Malz, F. Stinzing, P. Woldt, P. Birien, W. Dutty, J. Franz, H.
Hamann, E. R\"ossle, H. Schledermann, H. Schmitt, H.-J. Urban, R.A.
Eisenstein, D. Hertzog, W. Oelert, G. Sehl, B.E. Bonner, G.
Ericsson, T. Johansson, S. Ohlsson, W.H. Breunlich, and P. Pawlek,
Nucl. Phys. A {\bf 526}, 575 (1991).

\bibitem{Pomp}
S. Pomp, {\em Hyperon Polarisation in the Reaction
$\overline{p}^{12}C \rightarrow {\PgL}{\PagL}X$}, Ph. D. thesis,
Uppsala University (1999).

\bibitem{FAIR08}
I. Augustin, H.H. Gutbrod, D. Kr\"amer, K. Langanke, H. St\"ocker,
Fourth International Conference on Fission and Properties of
Neutron-Rich nuclei,Sanibel Island, Florida, 2007; arXiv:0804.0177v1
[hep-ph].

\bibitem{Panda}
${\overline{\rm P}}$ANDA Collaboration, Technical Progress Report
(GSI Darmstadt), pp. 1-383 (2005).

\bibitem{Hah56}
B. Hahn, D.G. Ravenhall, and R. Hofstadter, Phys. Rev. {\bf 101},
1131 (1956).

\bibitem{PDG}
Review of Particle Physics, Particle Data Group, J. Phys. G: Nucl.
Part. Phys. {\bf 33}, 1 (2006).

\bibitem{Sibirtsev99}
A. Sibirtsev, K. Tsushima, and A.W. Thomas, Eur. Phys.J. A{\bf 6},
351 (1999).

\bibitem{Lenske05}
H. Lenske and P. Kienle, Phys. Lett. B {\bf 647}, 82 (2007).

\bibitem{Moniz71}
E.J. Moniz, I. Sick, R.R. Whitney, J.R. Ficenec, R.D. Kephart, and
W.P. Trower, Phys. Rev. Lett. {\bf 26}, 445 (1971).

\bibitem{Wise93}
J.E. Wise, M.R. Braunstein, S. Høibråten, M.D. Kohler, B. J. Kriss,
J. Ouyang, R.J. Peterson, J. A. McGill, C.L. Morris, S.J. Seestrom,
R.M. Whitton, J.D. Zumbro, C.M. Edwards and A.L. Williams, Phys.
Rev. C {\bf 48}, 1840 (1993).

\bibitem{Kormanyos95}
C.M. Kormanyos, R.J. Peterson, J.R. Shepard, J.E. Wise, S. Bart,
R.E. Chrien, L. Lee, B.L. Clausen, J. Piekarewicz, M. B. Barakat,
E.V. Hungerford, R.A. Michael, K.H. Hicks, and T. Kishimoto, Phys.
Rev. C {\bf 51}, 669 (1995).

\bibitem{Fujii01}
Y. Fujii, O. Hashimoto, T. Nakagawa, Y. Sato, T. Takahashi, J. T.
Brack, C. J. Gelderloos, M. V. Keilman, R. J. Peterson, M. Itoh, H.
Sakaguchi, H. Takeda, K. Aoki, H. Hotchi, H. Noumi, Y. Ohta, H.
Outa, M. Sekimoto, M. Youn, S. Ajimura, T. Kishimoto, H. Bhang, H.
Park, and R. Sawafta, Phys. Rev. C {\bf 64}, 034608-1 (2001).

\bibitem{Tabakin91}
F. Tabakin, R.A. Eisenstein, Y.Lu, Phys. Rev. C {\bf 44}, 1749
(1991).

\bibitem{Barnes96}
P.D. Barnes, G. Franklin, B. Quinn, R. Schumacher, V. Zeps, N.
Hamann, W. Dutty, H. Fischer, J. Franz, E. R\"ossle, H. Schmitt, R.
Todenhagen, R. v. Frankenberg, K. Kilian, W. Oelert, K. R\"ohrich,
K. Sachs, T. Sefzick, M. Ziolkowski, R.A. Eisenstein, P.G. Harris,
D.W. Hertzog, S.A. Hughes, P.E. Reimer, R.L. Tayloe, W. Eyrich, R.
Geyer, M. Kirsch, R.A. Kraft, F. Stinzing, T. Johansson and S.
Ohlsson, Phys. Rev. C {\bf 54}, 2831 (1996).

\bibitem{Weber02}
H. Weber, E.L. Bratkovskaya, and H. St\"ocker, Phys. Rev. C {\bf
66}, 054903 (2002).

\bibitem{Yamazaki99}
T. Yamazaki and Y. Akaishi, Phys. Lett. B {\bf  453}, 1 (1999).

\bibitem{Tsu03}
K. Tsushima and F.C. Khanna, Phys. Lett. B {\bf 552}, 138 (2003).

\bibitem{Sai05}
K. Saito, K. Tsushima, and A.W. Thomas, Prog.Part.Nucl.Phys. {\bf
58}, 1 (2007).

\bibitem{Kai94}
A.B. Kaidalov and P.E. Volkovitsky, Z. Phys. C {\bf 63}, 517 (1994).

\bibitem{PANDA_PB}
S. Grape, Licenciate thesis, Uppsala University (2008),
arXiv:0805.0950v1; Physics Performance Report for ${\overline{\rm
P}}$ANDA (in preparation).

\bibitem{Cas02}
W. Cassing, E.L. Bratkovskaya, and O. Hansen, Nucl. Phys. A {\bf
707}, 224 (2002).
\end{thebibliography}
\end{document}